# THE IMPORTANCE OF THE ALGORITHMIC INFORMATION THEORY TO CONSTRUCT A POSSIBLE EXAMPLE WHERE NP ≠ P - II: AN IRREDUCIBLE SENTENCE

Rubens Viana Ramos

rubens@deti.ufc.br

*Department of Teleinformatic Engineering – Federal University of Ceara - DETI/UFC*

*C.P. 6007 – Campus do Pici - 60755-640 Fortaleza-Ce Brazil*

In this short communication it is discussed the relation between disentangled states and algorithmic information theory aiming to construct an irreducible sentence whose length increases in a non-polynomial way when the number of qubits increases.

One of the most enigmatic properties of quantum mechanics is the quantum entanglement. Since the entanglement has found very important applications in quantum communication and computing [1], new ways to determine if a quantum state is entangled or not is a crucial point. Several entanglement measures have been proposed [2-5]. Among these, the quantum relative entropy is of particular interest. It is based on the distance between the state whose entanglement one wishes to measure and the set of disentangled states. In order to calculate it, it is interesting to know the general form of an *n*-way disentangled state. It is obvious that, if a *n*-partite of qubit quantum state cannot be written in the form of a general *n*-partite disentangled stated, then it is entangled. For example, a bipartite state $\Gamma_{AB}$ has 2-way entanglement if

$$\Gamma_{AB} \neq \sum_i p_i \left( \rho_A^i \otimes \rho_B^i \right) \qquad (1)$$

$$\sum_i p_i = 1 \qquad (2)$$

where $\rho_A$ and $\rho_B$ are single-qubit states. On the other hand, a tripartite state $\Gamma_{ABC}$ has 3-way entanglement if

$$\Gamma_{ABC} \neq \sum_i p_i \left(\rho_A^i \otimes \rho_B^i \otimes \rho_C^i\right) + \sum_j r_j \left(\rho_A^j \otimes \Phi_{BC}^j\right) + \sum_l q_l \left(\rho_B^l \otimes \Phi_{AC}^l\right) + \sum_k t_k \left(\rho_C^k \otimes \Phi_{AB}^k\right) \quad (3)$$

$$\sum_i p_i + \sum_j r_j + \sum_l q_l + \sum_k t_k = 1 \quad (4)$$

where $\Phi_{AB}$, $\Phi_{BC}$ and $\Phi_{AC}$ are entangled bipartite states. When the number of qubits $n$ increases, the number of terms of the general $n$-way disentangled state grows very fast. In fact it is easy to show that for $N>4$ the number of terms is larger than $2^N$. Since any term of the general $n$-way disentangled state represents physically a different kind of quantum state in the meaning that the positions of the entanglement are different, the sentence cannot be reduced to a smaller one containing a lower number of terms. Hence, this work proposes the general $n$-way disentangled state as an irreducible sentence, in the same sense that $\Omega$ [6,7] proposed by Chaitin is an irreducible binary number. Since the number of terms of a general $n$-qubit disentangled state grows in a non-polynomial way, any problem where it is necessary and the number of qubits increases is a NP problem that can not be reduced to a P problem.